\documentclass[traditabstract,a4]{aa}  

\usepackage{epsfig}
\usepackage{graphicx}
\usepackage{epstopdf}
\usepackage{color}  
\usepackage{array}   
\usepackage{units}  
\usepackage[breaklinks, colorlinks, citecolor=blue]{hyperref}
\usepackage{natbib}  
\usepackage{multirow}   
\usepackage{amsmath,amssymb}  
\usepackage[english]{babel}
\usepackage[latin1]{inputenc}
\usepackage[T1]{fontenc}
\usepackage{longtable,lscape}
\usepackage{footnote}
\usepackage{txfonts}
\usepackage[toc,page]{appendix} 

\def\be{\begin{equation}}
\def\ee{\end{equation}}
\def\ba#1\ea{\begin{align}#1\end{align}}

\def\mr{\mathrm}

\def\dd{\mathrm{d}}

\def\dd{\mathrm{d}}

\begin{document}


\title{Combined analysis of galaxy cluster number count, thermal Sunyaev-Zel'dovich power spectrum, and bispectrum.} 

\author{G. Hurier\inst{1,2} \& F. Lacasa\inst{3,4}}
\titlerunning{Cosmological constraints from the tSZ effect}

\institute{$^{1}$Centro de Estudios de F\'isica del Cosmos de Arag\'on (CEFCA),Plaza de San Juan, 1, planta 2, E-44001, Teruel, Spain\\
$^{2}$Institut d'Astrophysique Spatiale, CNRS (UMR8617) Universit\'{e} Paris-Sud 11, B\^{a}timent 121, Orsay, France
\\
\email{ghurier@cefca.es} \\
$^{3}$ICTP - South American Institute for Fundamental Research, Rua Dr. Bento Teobaldo Ferraz 271, S$\tilde{\mathrm{a}}$o Paulo, SP Brazil.
\\
$^{4}$ D\'{e}partement de Physique Th\'{e}orique, Universit\'{e} de Gen\`{e}ve, 24 quai Ernest Ansermet, CH-1211 Gen\`{e}ve 4, Switzerland. \\
\email{fabien.lacasa@unige.ch} 
}

\date{Received /Accepted}
 
\abstract{
   The Sunyaev-Zel'dovich (SZ) effect is a powerful probe of the evolution of structures in the universe, and is thus highly sensitive to cosmological parameters $\sigma_8$ and $\Omega_m$, though its power is hampered by the current uncertainties on the cluster mass calibration.
In this analysis we revisit constraints on these cosmological parameters as well as the hydrostatic mass bias, by performing (i) a robust estimation of the tSZ power-spectrum, (ii) a complete modeling and analysis of the tSZ bispectrum, and (iii) a combined analysis of galaxy clusters number count, tSZ power spectrum, and tSZ bispectrum. From this analysis, we derive as final constraints $\sigma_8 = 0.79 \pm 0.02$, $\Omega_{\rm m} = 0.29 \pm 0.02$, and $(1-b) = 0.71 \pm 0.07$. These results favour a high value for the hydrostatic mass bias compared to numerical simulations and weak-lensing based estimations. They are furthermore consistent with both previous tSZ analyses, CMB derived cosmological parameters, and ancillary estimations of the hydrostatic mass bias.
}


\maketitle


\section{Introduction}
\label{sec:introduction}

Galaxy clusters are the largest structure in the universe that are gravitationally bound. They are consequently a tailored probe of the evolution of the universe, in particular the growth of structure with cosmic time.
For example, the number counts of galaxy clusters has been shown to tightly scale with cosmological parameters, using numerical simulations \citep{tin08,wat12}.
Using galaxy clusters abundance or correlation functions is now a mature activity to constrain cosmological parameters \citep{has13,planckszc,dah16}. Galaxy clusters can be observed through a large number of observational probes: over-density of galaxies \citep{wen12,roz14}, lensing of background galaxies \citep{hey12,erb13}, X-ray emission from the hot intra-clusters electrons \citep{boh01}, and the thermal Sunyaev-Zel'dovich (tSZ) effect from the same electron populations \citep{sun72}.

The main limitation to the cosmological use of galaxy clusters is the necessity to properly calibrate the relation between the cluster masses and the probe used to detect them \citep{planckszc}.
In order to solve this mass-observable problem, it is possible to combine different probes, so as to simultaneously constrain the cosmological parameters and the mass-observable relations.

In this context, the tSZ effect \citep{sun72} is a tailored mass proxy.
This effect is a distortion of the CMB black body radiation through inverse Compton scattering. CMB photons receive an average energy boost when scattering off hot (a few keV) ionized electrons of the intra-cluster medium \citep[see e.g.][for reviews]{bir99,car02}.
The intensity of the tSZ effect in a given direction on the sky is measured by the thermal Compton parameter, $y$, which is related to the electron density along the line of sight by:
\begin{equation}
y (\vec{n}) = \int n_{e} \frac{k_{\rm{B}} T_{\rm{e}}}{m_{\rm{e}} c^{2} } \, \sigma_{T} \  \mr{d}s
\label{comppar}
\end{equation}
where $\mr{d}s$ is the distance along the line-of-sight,  $n_{\rm{e}}$
and $T_{e}$ are respectively the electron number density and temperature.
In units of CMB temperature the contribution of the tSZ effect for a given observation frequency $\nu$ is
\begin{equation}
\frac{\Delta T_{\rm{CMB}}}{T_{\rm{CMB}} }= g(\nu) \ y.
\end{equation}
where, neglecting relativistic corrections, we have the frequency factor 
\begin{equation}
g(\nu) = \left[ x\coth \left(\frac{x}{2}\right) - 4 \right] \quad \mr{with} \quad x=\frac{h \nu}{k_{\rm{B}} \, T_{\rm{CMB}}}
\label{szspec}
\end{equation}
where $T_{\rm CMB}$~=~2.726$\pm$0.001~K, the tSZ effect is negative below 217~GHz and positive for higher frequencies.\\

Tight cosmological constraints have been obtained from the tSZ signal using cluster number counts \citep{planckszc}, the tSZ angular power spectrum \citep{planckszs}, or more recently the tSZ bispectrum \citep{geo15,planckszs}.
The present work aims at combining the constraining power from the cluster number counts, tSZ angular power spectrum and bispectrum, in order to set joint constraints on the cosmological parameters and the mass-observable relation.

The paper is organized as follows, section~\ref{seccl} details the modeling of the tSZ effect angular power spectrum, then section~\ref{secparcl} present a new approach to measure the tSZ angular power spectrum with a low level of contamination and the derived cosmological constraints. Then, section~\ref{secbl} presents the modeling of the tSZ bispectrum and section~\ref{secparbl} the derived cosmological constraints. Finally, in section~\ref{seccomb}, we present a combined analysis of all probes : cluster number count, tSZ angular power spectrum and bispectrum, and present the resulting cosmological and mass-observable relation constraints.


\section{Power spectra}
\label{seccl}
\subsection{General formalism}
The angular power spectrum reads
\begin{equation}
C_\ell = \frac{1}{2\ell +1} \sum_{m} |y_{\ell m}|^2,
\end{equation}
with $y_{\ell m}$ the harmonic coefficients of the tSZ map. 
In the context of large scale structure tracers, we model these correlations, assuming the following general expression
\begin{equation}
C_{\ell} = C^{{\rm 1h}}_\ell + C^{{\rm 2h}}_\ell,
\end{equation}
where $C^{{\rm 1h}}_\ell$ is the one-halo contribution and $C^{{\rm 2h}}_\ell$ is the two-halo contribution. These terms can be computed considering a halo model formalism. A key ingredient is the mass function, ${\dd n_\mr{h}}/{\dd M}$, which gives the abundance of dark matter halos depending on their mass and redshift. In this article we will be using the fitting formula from \citet{tin08}.\\

The one-halo/Poissonian term can be computed using the following ingredients: the Fourier transform of the normalized halo projected profiles in the tSZ map, the mass function, and the tSZ flux of the halo \citep[see e.g.][for a derivation of the tSZ auto-correlation angular power spectrum]{kom02}.
{\small
\begin{equation}
C_{\ell}^{{\rm 1h}} = 4 \pi \int {\rm d}z \ \frac{{\rm d}V}{{\rm d}z {\rm d}\Omega}\int{\rm d}M \ \frac{\dd n_\mr{h}}{\dd M} \, {Y}^2_{500} \, y^2_{\ell},
\end{equation}
}
where ${Y}_{500}$ ais the spherical tSZ halo flux in the $R_{500}$ radius. This flux depends on $M_{500}$ and $z$ and can be obtained with scaling relations. ${{\rm d}V}/{{\rm d}z {\rm d}\Omega}$ is the comoving volume element. 
The line-of-sight projected Fourier transform of the 3-D profile reads
\be
y_\ell(M_\mr{500},z) = \frac{4 \pi r_{\rm s}}{l^2_{\rm s}} \int_0^{\infty} {\rm d}x \, x^2 p(x) \frac{{\rm sin}(\ell x / \ell_{\rm s})}{\ell x / \ell_{\rm s}},
\ee 
 where $p(x)$ is the tSZ halo 3-D profile, $x = r/r_{\rm s}$, $\ell_{{\rm s}} = D_{\rm ang}(z)/r_{\rm s}$, and $r_{\rm s}$ is the scale radius of the halo.\\

The two-halo term is due to large scale fluctuations of the dark matter field, that induce correlations in the halo distribution over the sky.
It can be computed as \citep[see e.g.][]{kom99,die04,tab11}
\begin{align}
C_{\ell}^{{\rm 2h}} = 4 \pi \int {\rm d}z \frac{{\rm d}V}{{\rm d}z{\rm d}\Omega}&\left(\int{\rm d}M \frac{\dd n_\mr{h}}{\dd M} {Y}_{500} y_{\ell} b(M,z)\right)\\ \nonumber
&\times \left(\int{\rm d}M \frac{\dd n_\mr{h}}{\dd M} {Y}_{500} y_{\ell} b(M,z)\right) P_m(k_\ell,z)
\end{align}
with $k_\ell = (\ell+1/2)/r(z)$, $b(M,z)$ the linear bias, that relates the matter power spectrum, $P_m(k,z)$, to the power spectrum of the cluster distribution. 
Following \citet{mo96,kom99} we adopt 
$$b(M,z)=1+(\nu^2(M,z)-1)/\delta_c(z),$$
with $\nu(M,z) = \delta_c(z)/\left[D_g(z) \sigma(M)\right]$, $D_g(z)$ is the linear growth factor and $\delta_c(z)$ is the over-density threshold for spherical collapse.\\

\begin{figure}[!th]
\begin{center}
\includegraphics[scale=0.2]{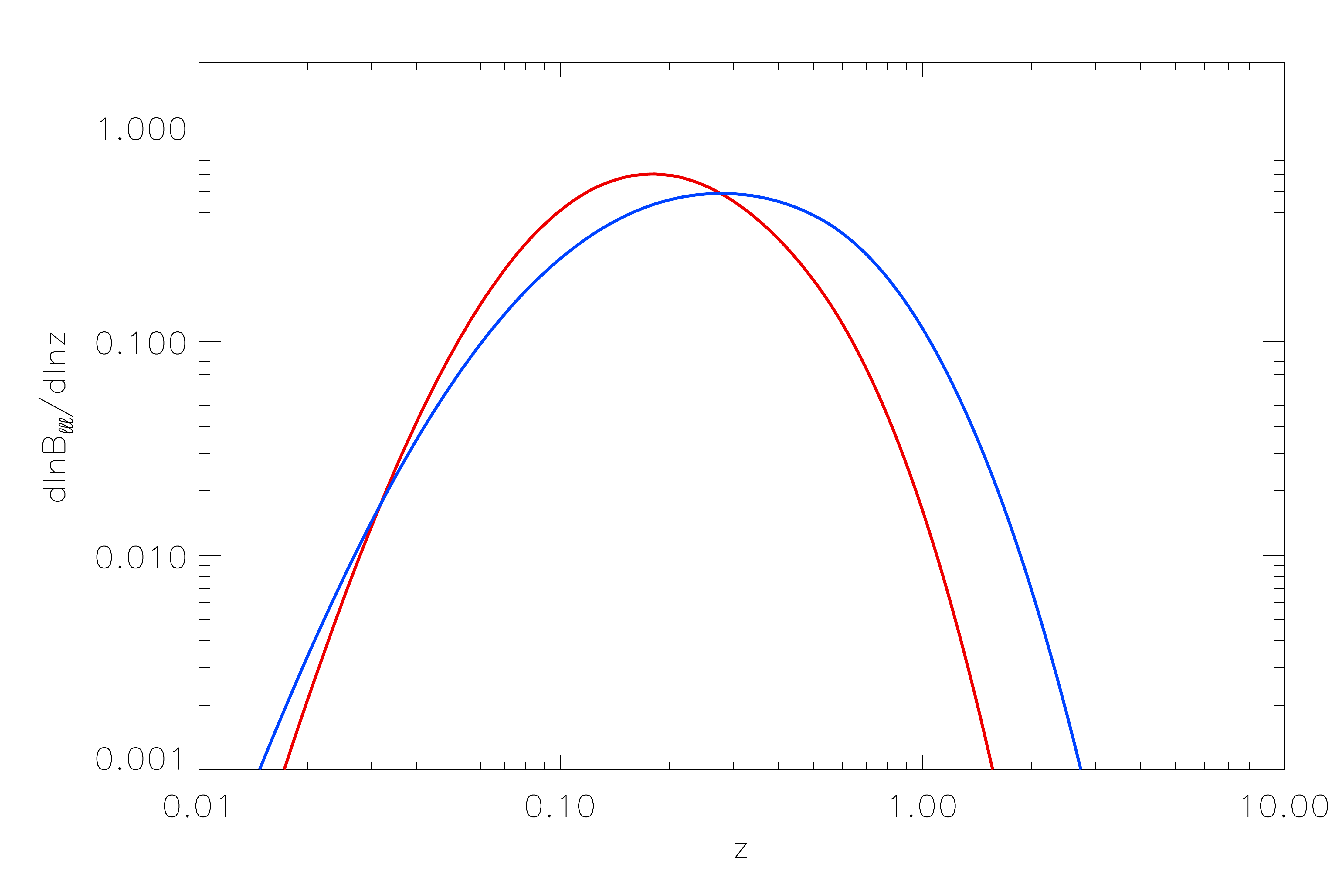}
\caption{Power density at $\ell = 500$ as a function of the redshift for the tSZ power spectrum in dark blue and for the bispectrum in red.}.
\label{figdz}
\end{center}
\end{figure}

\begin{figure}[!th]
\begin{center}
\includegraphics[scale=0.2]{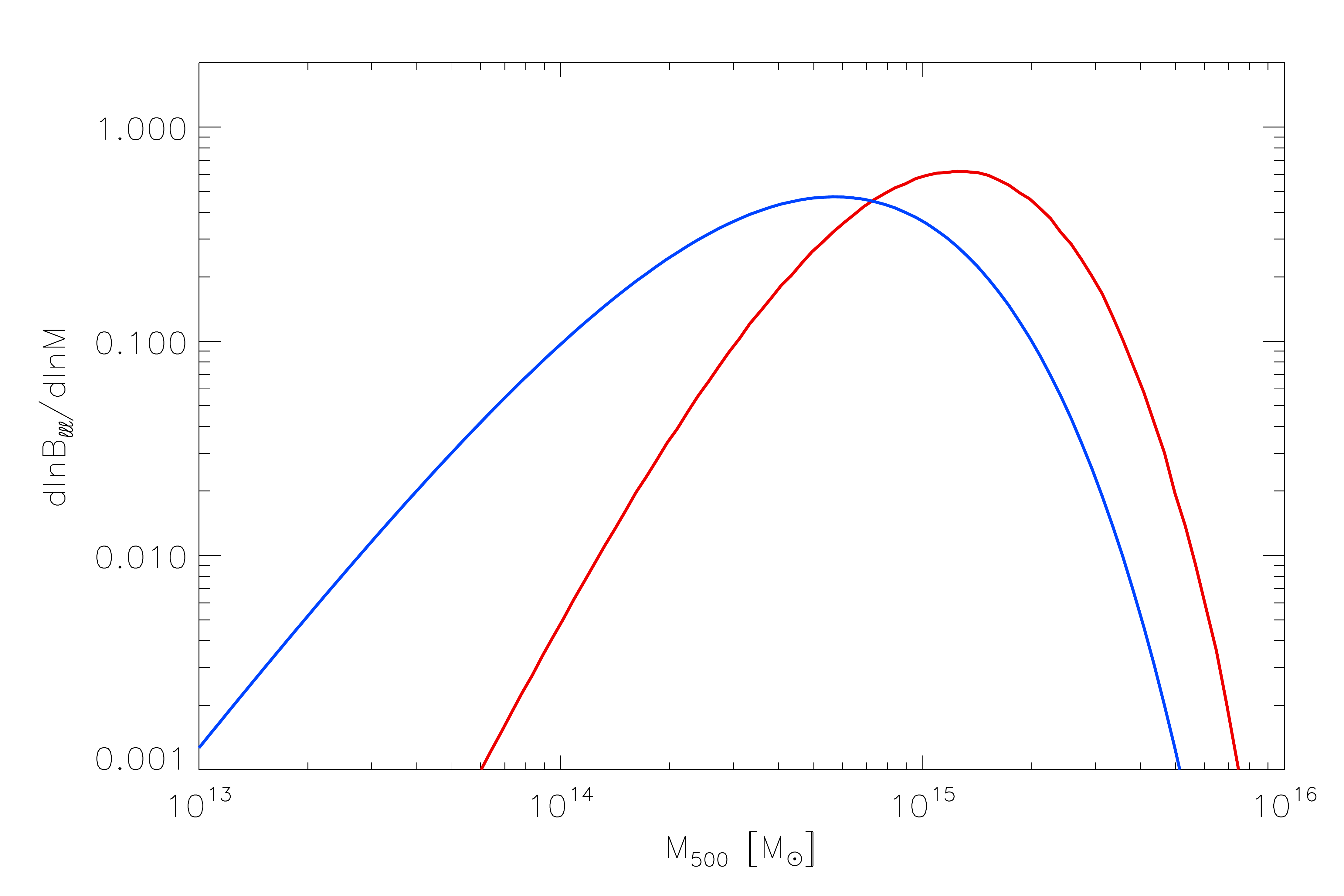}
\caption{Power density at $\ell = 500$ as a function of the mass, $M_{500}$, of dark matter halos for the tSZ power spectrum in dark blue and for the bispectrum in red.}.
\label{figdm}
\end{center}
\end{figure}

On Fig.~\ref{figdz}, we present the power distribution as a function of the redshift, showing that the tSZ power spectrum is dominated by objects at $z \leq 1$. 
The figure~\ref{figdm} presents the same power distribution as a function of the galaxy cluster masses. The tSZ power spectrum is dominated by halos with $M_{500} > 10^{14}\, {\rm M}_\odot$, but also receives contribution from smaller halos down to $M_{500} = 10^{13}\, {\rm M}_\odot$.
The power distributions for the tSZ power spectrum are presented as solid blue lines, solid red lines show the same distributions for the tSZ bispectrum and will be discussed in Sect.~\ref{secbl}.

\subsection{The tSZ scaling relation}
\label{secscal}

A key step in the modeling of the tSZ effect is to relate the mass, $M_{500}$, and the redshift, $z$, of a given cluster to its tSZ flux, $Y_{500}$. This relation has to be calibrated on a representative sample of galaxy clusters, with careful propagation of statistical and systematic uncertainties. In this work we use the $M_{500}-Y_{500}$ scaling laws presented in \citet{planckszc}, 
\begin{equation}
E^{-\beta_{\rm sz}}(z) \left[ \frac{D^2_{ang}(z) {Y}_{500}}{10^{-4}\,{\rm Mpc}^2} \right] = Y_\star \left[ \frac{h}{0.7} \right]^{-2+\alpha_{\rm sz}} \left[ \frac{(1-b) M_{500}}{6 \times 10^{14}\,{\rm M_{\odot}}}\right]^{\alpha_{\rm sz}},
\label{szlaw}
\end{equation}
with $E(z) = \Omega_{\rm m}(1+z)^3 + \Omega_{\Lambda}$. The coefficients $Y_\star$, $\alpha_{\rm sz}$, and $\beta_{\rm sz}$ from \citet{planckszc}, are given in Table~\ref{tabscal}. We used $b=0.2$ for the bias between the X-ray estimated mass and the effective mass of the clusters (hydrostatic mass bias).

\subsection{Pressure and density profiles}
\label{secprof}

The tSZ effect is directly proportional to the pressure integrated along the line of sight. In this work, we model the galaxy cluster pressure profile by a generalized Navarro Frenk and White \citep[GNFW,][]{nav97,nag07} profile of the form
\begin{equation}
{\mathbb P}(r) = \frac{P_0}{\left(c_{500} r\right)^\gamma \left[1 + (c_{500} r)^\alpha \right]^{(\beta-\gamma)/\alpha}}.
\end{equation}
For the parameters $c_{500}$, $\alpha$, $\beta$, and $\gamma$, we used the best-fitting values from \citet{arn10} presented in Table.~\ref{tabscal}. The absolute normalization of the profile $P_0$ is set assuming the scaling laws $Y_{500}-M_{500}$ presented in Sect.~\ref{secscal}.\\

\begin{table}
\center
\caption{Scaling-law parameters and error budget for both $Y_{500}-M_{500}$ \citep{planckszc}, $K_{500}-M_{500}$, and $Y_{500}-T_{500}$ \citep{planckszc} relations}
\begin{tabular}{|cc|cc|}
\hline
\multicolumn{2}{|c|}{$M_{500}-Y_{500}$} & \multicolumn{2}{|c|}{$M_{500}-T_{500}$} \\
\hline
${\rm log}\,Y_\star$ & -0.19 $\pm$ 0.02 & ${\rm log}\, T_\star$ & -4.27 $\pm$ 0.02   \\
$\alpha_{\rm sz}$ & 1.79 $\pm$ 0.08 & $\alpha_{\rm T}$ & 2.85 $\pm$ 0.18  \\
$\beta_{\rm sz}$ & 0.66 $\pm$ 0.50 &  $\beta_{\rm T}$ & 1 \\
$\sigma_{{\rm log}\, Y}$ & 0.075 $\pm$ 0.010 & $\sigma_{{\rm log}\, T}$ & 0.14 $\pm$ 0.02 \\
\hline
\end{tabular}
\label{tabscal}
\end{table}


\begin{figure*}[!th]
\begin{center}
\includegraphics[scale=0.4]{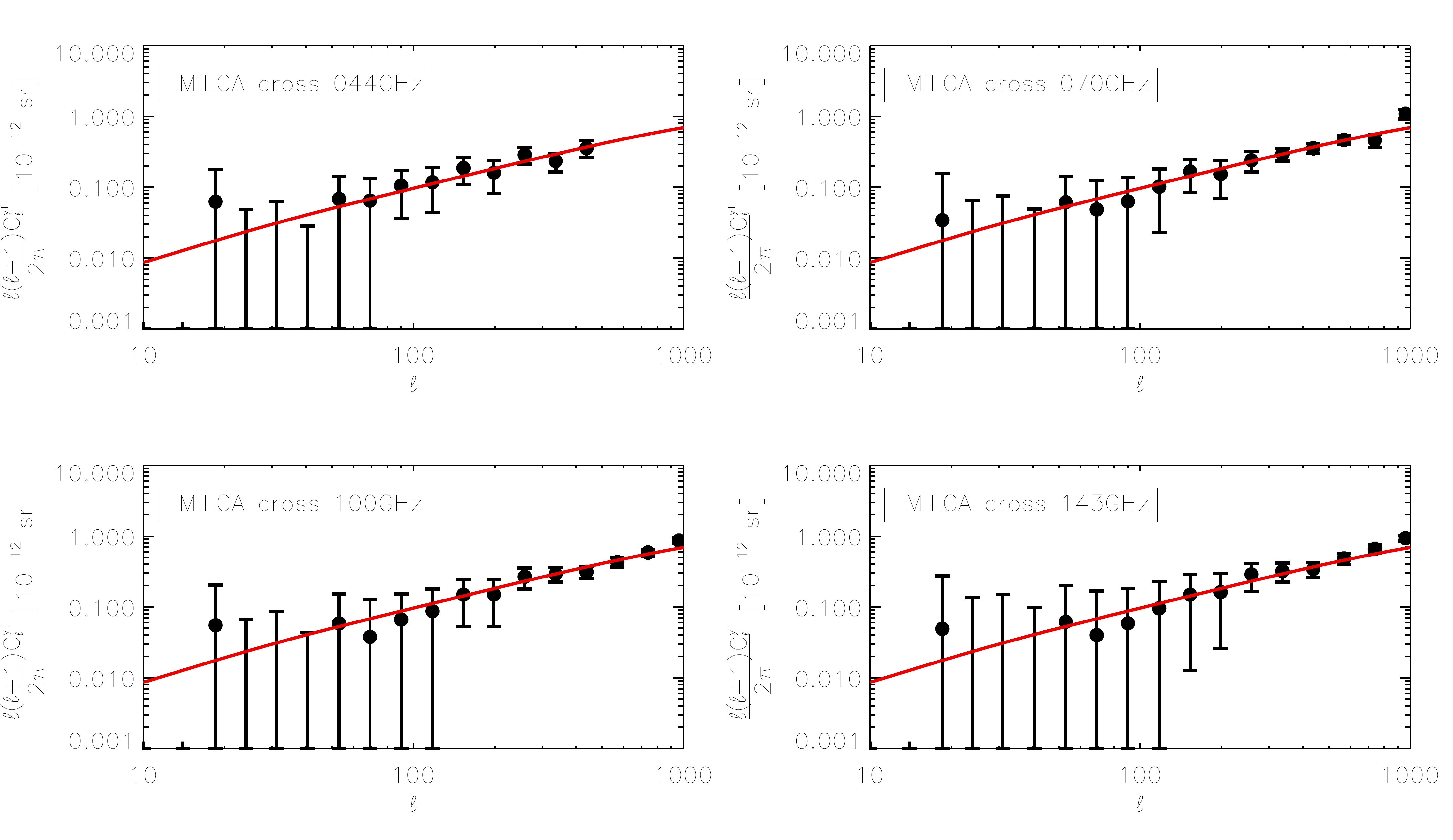}
\caption{tSZ power spectra reconstructed by cross-correlating Planck intensity map and MILCA $y$-map (black sample). From left to right and top to bottom: at 44, 70, 100, and 143 GHz. The best fitting model is presented by the solid red lines. We display the spectra in Compton parameter units, $\widehat{C}_\ell^{y\nu}/g(\nu)$, to ease the comparison between cross-spectra.}.
\label{figcl}
\end{center}
\end{figure*}


\section{Cosmological constraints from the tSZ power spectrum}
 
 \label{secparcl}
 
In this section we revisit the cosmological constraints derived from the tSZ angular power spectrum by \cite{planckszs}. The construction of tSZ maps \citep[e.g.,][]{rem11,hur13,gmca} is now a mature subject, however they still often suffer from significant contamination by different foregrounds, in particular by extra-galactic radio-sources and by the cosmic infra-red background (CIB) \citep{planckszs}.
Previous works have handled this sources of contamination using a complete modeling and propagation of the CIB component from the frequency maps to the tSZ power-spectrum \citep{planckszs, planckSZCIB}.
We present here a new approach to disentangle the tSZ signal from the contribution produced by other astrophysical components. Then, we use these measurements to put constraints on cosmological parameters $\sigma_8$ and $\Omega_b$.
 
\subsection{Measurement}

We start by computing a tSZ $y$-map at 7 arcmin resolution using the MILCA procedure \citep{hur13} from the nine Planck frequencies (30, 44, 70, 100, 143, 217, 353, 545 and 957 GHz) \citep{planckMIS}. 
By construction the instrumental noise contribution in MILCA maps is correlated with the instrumental noise in Planck intensity maps. In order to cancel this noise contribution, we use the so-called jacknife methodology : given two independent sub-datasets containing half of the the Planck data, we compute $C_\ell^{y\nu}$ as
\begin{align}\label{Eq:Clynu}
C_\ell^{y\nu} = \frac{1}{2}\left( C_\ell^{y_1\nu_2} + C_\ell^{y_2\nu_1} \right),
\end{align}
where $y_i$ is the MILCA tSZ map coming from sub-dataset $i$, and $\nu_i$ is the corresponding Planck intensity map at frequency $\nu$. We precise that the two MILCA maps, have been built using the same linear combination determined on the complete dataset and then applied to the two sub-datasets.
We thus obtain nine angular cross power spectra, $C_\ell^{y\nu}$, each spectra being corrected for the beam and the mask effect following \cite{tri03}.
In these cross spectra, the tSZ contribution follows the $g(\nu)$ spectral energy distribution. It allows to separate the tSZ contribution from other residual emissions in the MILCA map (mainly radio sources and CIB contributions).\\
The variance of $C_\ell^{y\nu}$ spectra reads,
\begin{align}
<\left(C_\ell^{y\nu}\right)^2> =& \frac{
\left(C_\ell^{y_1\nu_2}\right)^2 + \left(C_\ell^{y_2\nu_1}\right)^2 + 2C_\ell^{y_1\nu_2}C_\ell^{y_2\nu_1}}{4(2\ell +1) f_{\rm sky}} \nonumber\\ 
&+ \frac{C_\ell^{\nu_1 \nu_1}C_\ell^{y_2y_2} + C_\ell^{\nu_2 \nu_2}C_\ell^{y_1y_1} + 2 C_\ell^{\nu_1 \nu_2}C_\ell^{y_1y_2}
}{4(2\ell +1) f_{\rm sky}},
\label{var}
\end{align}
where, $C_\ell^{\nu \nu}$ is the auto correlation of the Planck intensity map at frequency $\nu$, $C_\ell^{yy}$ is the auto-correlation of the tSZ Compton parameter map, and $f_{\rm sky}$ the covered sky fraction. In the present analysis we use the same mask as used in \citet{planckszs}.\\
The noise is dominated by the second term in Eq.~\ref{var}, and especially the contribution from astrophysical emissions.
Thus, the cross-spectra are highly correlated from frequency to frequency. The covariance can be computed as,
\begin{align}
<C_\ell^{y\nu}C_\ell^{y\nu'}> =& \frac{
C_\ell^{y_1\nu_2}C_\ell^{y_1\nu'_2} + C_\ell^{y_2\nu_1}C_\ell^{y_2\nu'_1} + C_\ell^{y_1\nu'_2}C_\ell^{y_2\nu_1} + C_\ell^{y_1\nu_2}C_\ell^{y_2\nu'_1}}{4(2\ell +1) f_{\rm sky}} \nonumber\\ 
&+ \frac{C_\ell^{\nu_1 \nu'_1}C_\ell^{y_2y_2} + C_\ell^{\nu_2 \nu'_2}C_\ell^{y_1y_1} + C_\ell^{\nu_1 \nu'_2}C_\ell^{y_1y_2} + C_\ell^{\nu'_1 \nu_2}C_\ell^{y_1y_2}
}{4(2\ell +1) f_{\rm sky}},
\label{cov}
\end{align}
We then perform two additional steps in order to clean the cross-spectra from non-tSZ signal. 
First, we use the cross-spectra at 217 GHz, $C_\ell^{y,217}$, to clean for CMB induced variance \footnote{variance due to the possibility of chance correlations between the CMB and tSZ map} in the power spectra estimates,
\begin{align}
\tilde{C}_\ell^{y\nu} =  \frac{g(\nu)}{g(\nu)-g(217)}\left( C_\ell^{y\nu} - C_\ell^{y,217} \right),
\end{align}
Additionally, we use the 857 GHz cross-spectra, $C_\ell^{y,857}$, where the tSZ power spectrum is expected to be negligible to clean for CIB contribution at first order, 
\begin{align}
\widehat{C}_\ell^{y\nu} = \frac{g(\nu)}{g(\nu)-\alpha_{\nu,857}g(857)} \left( \tilde{C}_\ell^{y\nu} - \alpha_{\nu,857}\tilde{C}_\ell^{y,857} \right),
\end{align}
where,
\begin{align} 
\alpha_{\nu,857} = \frac{<\ell\tilde{C}_\ell^{\nu,857}>_{\ell \in [1000-2000]}}{<\ell \tilde{C}_\ell^{857,857}>_{\ell \in [1000-2000]}},
\end{align}
is an estimation of the infra-red contamination SED with $\tilde{C}_\ell^{\nu,857} = {C}_\ell^{\nu,857} - {C}_\ell^{217,857}$.\\

Such cleaning procedure presents limitation as described in \citet{hur14}, however at low-frequency (< 217 GHz), the CIB intensity is faint, and thus will not bias significantly the amplitude of $\widehat{C}_\ell^{y\nu}$, except at 353 and 545 GHz. These two frequencies have thus been excluded from the analysis.
 We also excluded the 30 GHz cross-spectra due to its poor angular resolution ($\simeq 30$') and the contamination from galactic and extra-galactic radio sources.\\
The two cleaning steps presented above are completely linear, thus we can propagate the $C_\ell^{y\nu}$ covariance matrix through the linear cleaning operation to determine the $\widehat{C}_\ell^{y\nu}$ covariance matrix.

On Fig.~\ref{figcl}, we present the derived cross-power spectra at 44, 70, 100, and 143 GHz. The global best fitting model on the four spectra is presented as a solid red line. We observe that the spectra from 44 to 143 GHz are consistent with a tSZ spectral behavior, and do not present evidence of a significant contamination from other astrophysical emissions. Such contamination would indeed appear as a frequency dependent bias on the spectra.
We stress that, due to the cleaning process and astrophysical sources of noise, the four cross-spectra present highly correlated uncertainties. Thus, the main advantage to have access to this four power spectra is to control and access contamination by other astrophysical sources.

\subsection{Cosmological constraints}

\begin{figure}[!th]
\begin{center}
\includegraphics[scale=0.2]{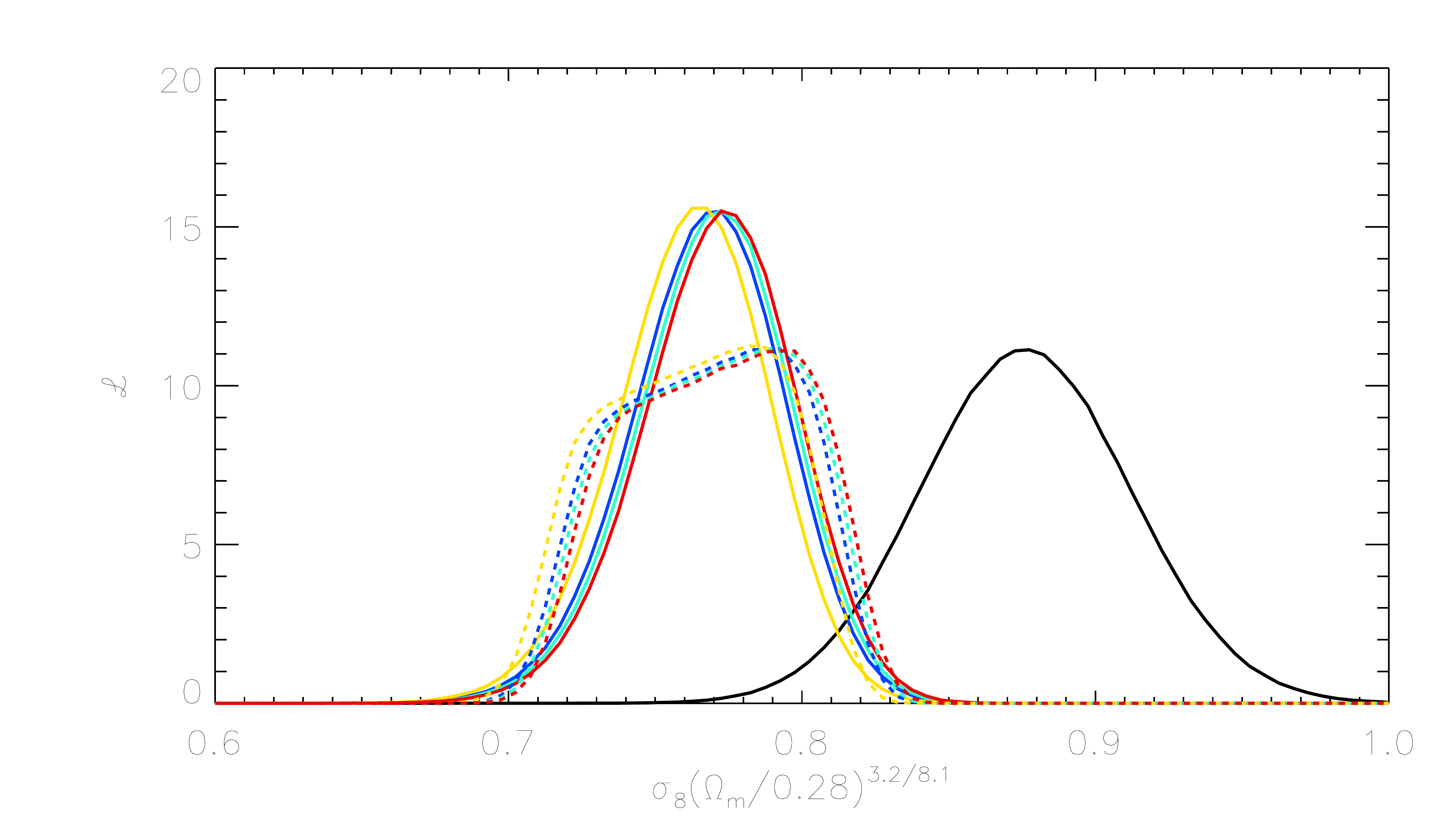}
\caption{Cosmological parameters likelihood function derived from MILCA-Planck maps cross-correlation, at 44 (dark blue), 70 (green), 100 (yellow), and 143 GHz (red). Solid lines presents the constraints assuming a gaussian prior on $(1-b) = 0.8 \pm 0.05$, the dashed lines assume a flat prior $0.7 < (1-b) < 0.9$. The solid black line presents the constraints provided by the CMB angular power spectrum}.
\label{figclc1}
\end{center}
\end{figure}

We fit for cosmological parameters using each of the cross spectra from 44 to 143 GHz. As shown in a previous analysis \citep{hur14} the shape of the angular power spectrum is essentially sensitive to the $Y-M$ mapping and does not significantly depend on cosmological parameters for the angular scales observed with Planck ($\ell < 2000$). Cosmological parameters only affect the overall normalization of the tSZ angular power spectrum. 
Consequently, there is a degeneracy between parameters $\sigma_8$ and $\Omega_m$, and we can only fit for the amplitude $S_8 = \sigma_8*(\Omega_{\rm m}/0.28)^{3.2/8.1}$.
The figure~\ref{figclc1}, shows the likelihood function for the four cross-spectra we used, for two possible priors on the bias of the hydrostatic mass : (i) a gaussian prior $(1-b) = 0.8 \pm 0.05$, and (ii) a flat prior $0.7 < (1-b) < 0.9$. For illustration, these constraints are compared with the cosmological constraints derived from the CMB power spectrum by \cite{PlanckPAR}. 
We find that the uncertainties on $S_8$ are completely dominated by the uncertainties on the $Y-M$ mapping.\\
We summarize in table.~\ref{cltab} the constraints derived from each spectra. Combining all spectra we obtain $S_8 = 0.77 \pm 0.02$, consistently with previous tSZ analysis using the same assumptions on the $Y-M$ mapping. In that case the CMB best fitting  parameters can be recovered for an hydrostatic mass bias of $(1-b) \simeq 0.6$ \citep[see,][for a detailed discussion]{planckszc}.\\

\begin{table}
\center
\caption{Cosmological constraints derived from the tSZ effect reconstructed through the cross-correlation angular power spectrum between MILCA $y$-map and Planck intensity maps from 44 to 143 GHz.}
\begin{tabular}{|c|c|c|}
\hline
$\nu$ & $\sigma_8(\Omega_{\rm m}/0.28)^{0.395}$ & $\sigma_8(\Omega_{\rm m}/0.28)^{0.395}((1-b)/0.8)^{0.442}$ \\
\hline
44 GHz & $0.770 \pm 0.024$ & $0.770 \pm 0.015$ \\
70 GHz & $0.770 \pm 0.022$ & $0.770 \pm 0.008$ \\
100 GHz & $0.764 \pm 0.022$ & $0.764 \pm 0.009$ \\
143 GHz & $0.773 \pm 0.023$ & $0.773 \pm 0.011$ \\
\hline
All & $0.770 \pm 0.021$ & $0.770 \pm 0.007$ \\
\hline
\end{tabular}
\label{cltab}
\end{table}



\section{Bispectrum modeling}

\label{secbl}

The angular tSZ equilateral bispectrum is a projection of the corresponding 3D equilateral bispectrum, with Limber's approximation giving:
\be
b_{\ell\ell\ell}^\mr{tSZ} = \int \dd z \frac{\dd V}{\dd z \dd \Omega} \; B_\mr{tSZ}(k_{\ell}, z)
\ee
with $\dd V = r^2(z) \, \frac{\dd r}{\dd z}$ the comoving volume per unit solid angle.\\
The 3D bispectrum is composed of 1-halo, 2-halo and 3-halo terms, with \citep{Lacasa2014-phd}:
\ba
B_\mr{1h}(k_\ell, z) &= \int \dd M \, \frac{\dd n_\mr{h}}{\dd M} \, \left(Y_\mr{500}(M,z) \ y_\ell(M,z)\right)^3 \\
\nonumber B_\mr{2h}(k_\ell, z) &=  3 \int \dd M_\texttt{ab} \left.\frac{\dd n_\mr{h}}{\dd M}\right|_{M_\texttt{a}} \left.\frac{\dd n_\mr{h}}{\dd M}\right|_{M_\texttt{b}} \, \Big( Y_\mr{500}(M_\texttt{a},z) \ y_\ell(M_\texttt{a},z) \Big)^2 \\
&\qquad \times Y_\mr{500}(M_\texttt{b},z) \ y_\ell(M_\texttt{b},z) \ P_\mr{halo}(k_\ell | M_\texttt{a}, M_\texttt{b}, z) \\
\nonumber B_\mr{3h}(k_\ell, z) &= \int \dd M_{123} \, \left(\prod_{i=1,2,3} \left.\frac{\dd n_\mr{h}}{\dd M}\right|_{M_i} \ Y_\mr{500}(M_i,z) \ y_\ell(M_i,z) \right) \\
& \qquad \times B_\mr{halo}(k_\ell | M_{123},z)
\ea
We take the halo power spectrum and bispectrum at the lowest order in bias and perturbation theory (tree-level):
\ba
P_\mr{halo}(k | M_\texttt{a}, M_\texttt{b}, z) & = b_1(M_\texttt{a},z) \, b_1(M_\texttt{b},z) \ P_m(k,z) \\
B_\mr{halo}(k | M_{123},z) &= B^\mr{2PT}_\mr{halo} + B^\mr{b2}_\mr{halo} \\
\nonumber B^\mr{2PT}_\mr{halo}(k | M_{123},z) &= 6 \ b_1(M_1,z) \, b_1(M_2,z) \, b_1(M_3,z)\\
& \quad \times F_\mr{equi} \ P_m(k,z)^2 \quad \mr{with} \quad F_\mr{equi} = \frac{2}{7}\\
\nonumber B^\mr{b2}_\mr{halo}(k | M_{123},z) &= \Big(b_1(M_1,z) \, b_1(M_2,z) \, b_2(M_3,z) \\
& \quad + 2 \ \mr{perm.}\Big) \times P_m(k,z)^2
\ea

The red lines on Fig.~\ref{figdz}~and~\ref{figdm} show the power density of the tSZ bispectrum for equilateral triangle at $\ell = 500$ as a function of redshift and mass respectively.
We observe that the tSZ bispectrum power density as a function of redshift is similar to the tSZ power spectrum power density, however favoring slightly lower redshift objects. The power density as a function of the mass shows that the tSZ bispectrum is dominated by objects with $M_{500} \simeq 10^{15}\, {\rm M}_\odot$, and favors more massive halos compared to the tSZ angular power spectrum.
These two findings are consistent with our expectations. Indeed the bispectrum, being a higher order quantity, is sensitive to more luminous objects than the power spectrum. Consequently, the bispectrum is dominated by a fewer number of objects and presents a higher sensitivity to the cosmic variance.


\section{Cosmological constraints from the tSZ bispectrum}

\label{secparbl}

\subsection{Measurement}

To measure the bispectrum of a $y(\vec{n})$ map, we use the following estimator
\begin{align}
b_{\ell_1,\ell_2\ell_3} &= \int \frac{\dd^2 \vec{n}}{4\pi} \ T_{\ell_1}(\vec{n}) \, T_{\ell_2}(\vec{n}) \, T_{\ell_3}(\vec{n})
\end{align}
where $T_{\ell}$ is the so-called scale maps that only contain harmonic coefficients of order $\ell$, i.e. $T_{\ell}(\vec{n})=\sum_m y_{\ell m} Y_{\ell m}(\vec{n})$ with $y_{\ell m}$ the harmonic coefficients of the Compton parameter map.\\
We refer to \citet{planckszs} for a more detailed description of the bispectrum estimation.

\begin{figure}[!th]
\begin{center}
\includegraphics[scale=0.2]{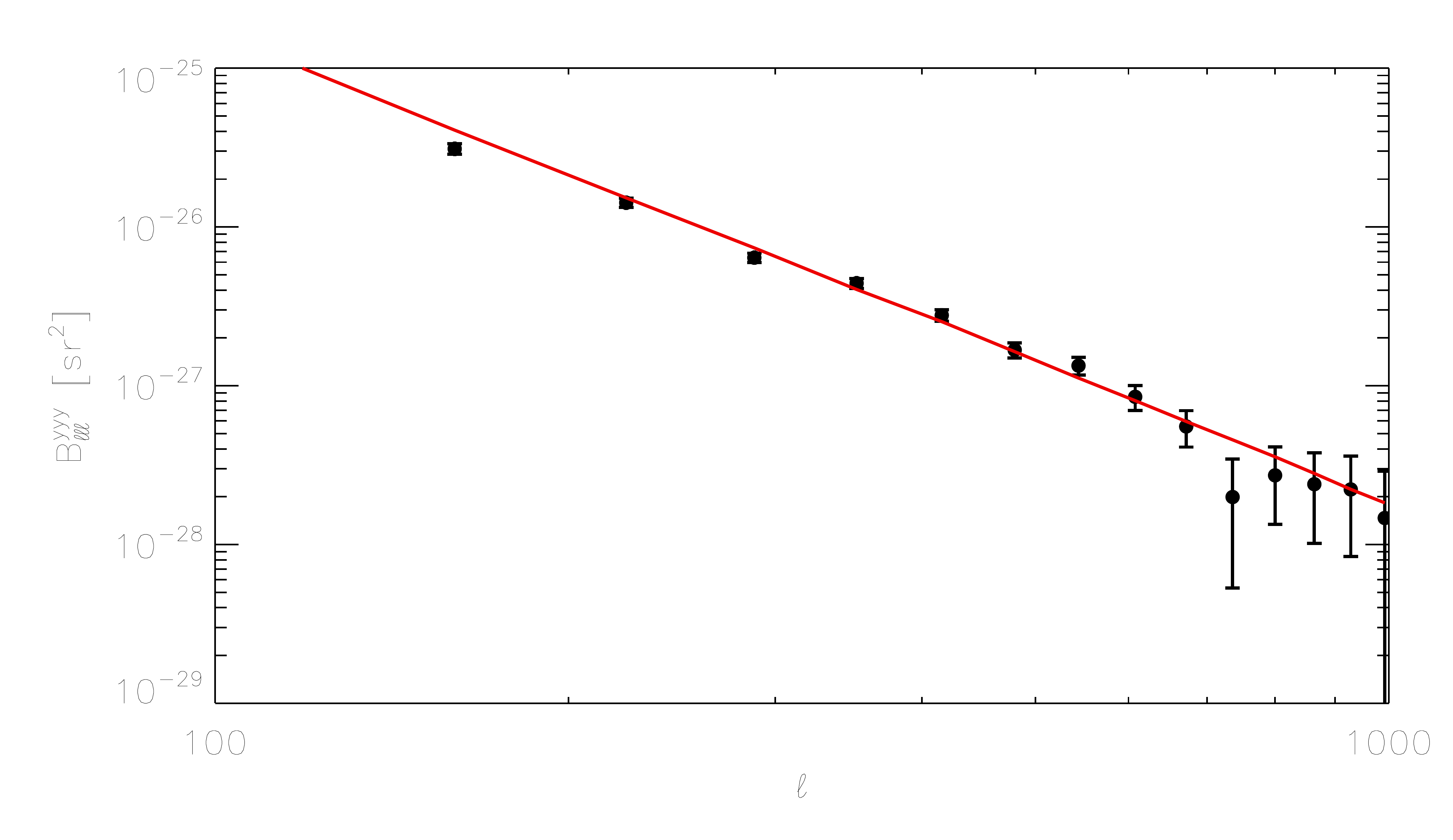}
\caption{tSZ bispectrum measured on the MILCA $y$-map (black sample). The best fitting model is shown as a red solid line.}.
\label{figbl}
\end{center}
\end{figure}

The uncertainties on the bispectrum are usually estimated under the weak non-gaussian limit and can be expressed as
\begin{align}
\label{eqvar}
<b_{\ell_1\ell_2\ell_3},b_{\ell'_1\ell'_2\ell'_3}> =& \frac{C_{\ell_1}C_{\ell_2}C_{\ell_3}}{N_{\ell_1,\ell_2,\ell_3}}\ \Big(\delta_{\ell_1 \ell'_1}\delta_{\ell_2 \ell'_2}\delta_{\ell_3 \ell'_3} \nonumber \\
&+\delta_{\ell_1 \ell'_1}\delta_{\ell_2 \ell'_3}\delta_{\ell_3 \ell'_2} +\delta_{\ell_1 \ell'_2}\delta_{\ell_2 \ell'_1}\delta_{\ell_3 \ell'_3} \nonumber \\
&+\delta_{\ell_1 \ell'_2}\delta_{\ell_2 \ell'_3}\delta_{\ell_3 \ell'_1} +\delta_{\ell_1 \ell'_3}\delta_{\ell_2 \ell'_2}\delta_{\ell_3 \ell'_1}\nonumber \\
&+\delta_{\ell_1 \ell'_3}\delta_{\ell_2 \ell'_1}\delta_{\ell_3 \ell'_2} \Big) ,
\end{align}
with $N_{\ell_1 \ell_2 \ell_3}$, being the number of modes for the ($\ell_1, \ell_2, \ell_3$) triangle.\\

In order to estimate the total uncertainty level in the bispectrum, we produced 100 tSZ simulated maps, using a Poissonian sampling of the cluster mass function and putting the corresponding halos randomly in the sky with tSZ fluxes following a log-normal distribution consistent with the tSZ $Y-M$ scaling relation.
The resulting total covariance is visible as the upper right panel of Fig.\ref{ccor}.


We note that the MILCA tSZ map is contaminated by non-gaussian astrophysical component that could in principle bias the measured tSZ bispectrum. First, a contamination by radio sources would appear as a negative contribution in the measured bispectrum at high-$\ell$. We avoided this contamination by applying aggressive radio source masks, and indeed find no trace of it in the measured bispectrum. Second, contamination by the Cosmic Infrared Background (CIB) is more delicate, as it cannot be masked and it produces a power spectrum similar to the tSZ one. Nevertheless, the CIB contribution in the $y$-map is essentially high-$z$ CIB \citep{planckSZCIB} and consequently is near-gaussian, i.e. with a low amplitude bispectrum. Furthermore, the (weak) bispectrum produced by the CIB is significantly less steep than the tSZ bispectrum \citep[e.g.,][]{Lacasa2014}, and would thus appear at high multipoles, which we do not see in our measurements.

On Figure~\ref{figbl}, we present the measured tSZ bispectrum on the MILCA tSZ-map (black sample) and the best fitting model (solid red line) for the equilateral configuration of the bispectrum. Uncertainties displayed on figure~\ref{figbl} only accounts for uncertainties induced by the instrumental noise and CIB-leakage. They do not include cosmic variance, that dominates the error budget at low-$\ell$.

\subsection{Cosmological constraints}

\begin{figure}[!th]
\begin{center}
\includegraphics[scale=0.2]{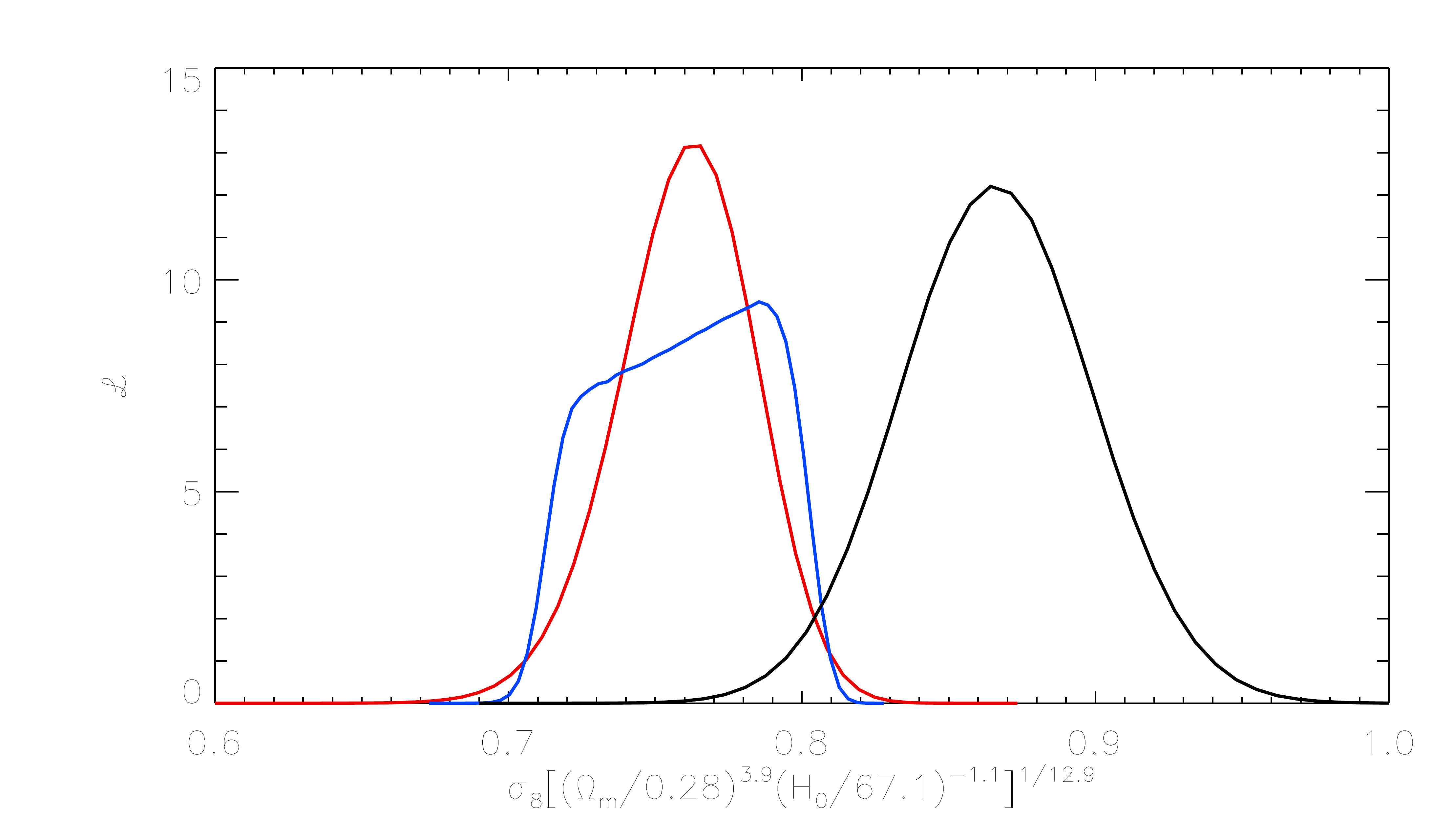}
\caption{Cosmological parameters likelihood function derived from MILCA $y$-map bispectrum. The red line shows the likelihood function assuming a gaussian prior $(1-b) = 0.8 \pm 0.05$ and the blue line a flat prior $0.7 < (1-b) < 0.9$. The black line shows the likelihood function derived from CMB angular power spectrum constraints.}.
\label{figblc}
\end{center}
\end{figure}

Figure~\ref{figblc} presents the cosmological constraints derived from the tSZ bispectrum for two cases: (i) assuming a gaussian prior of $(1-b) = 0.8 \pm 0.05$ and (ii) a flat prior $0.7 < (1-b) < 0.9$ for the bias on hydrostatic mass.
We derive $\sigma_8 \left[ \left(\Omega_{\rm m}/0.28\right)^{3.9}  \left(H_{0}/67.1\right)^{-1.1}\right]^{1/12.9} = 0.765 \pm 0.025$. 
This constraint is consistent with previous work \citep{planckszs} and with our derivation of cosmological parameters from the tSZ angular power spectrum in section~\ref{secparcl}.


\section{Combined analysis of number counts, power spectrum, and bispectrum}
\label{seccomb}
In this section, we combined our measurement of the tSZ power spectrum (from 44 to 143 GHz) and tSZ bispectrum with cluster number count analysis using the Planck cosmology sample from \citet{planckszc}.

\subsection{Covariance matrix}

\begin{figure}[!th]
\begin{center}
\includegraphics[scale=0.27]{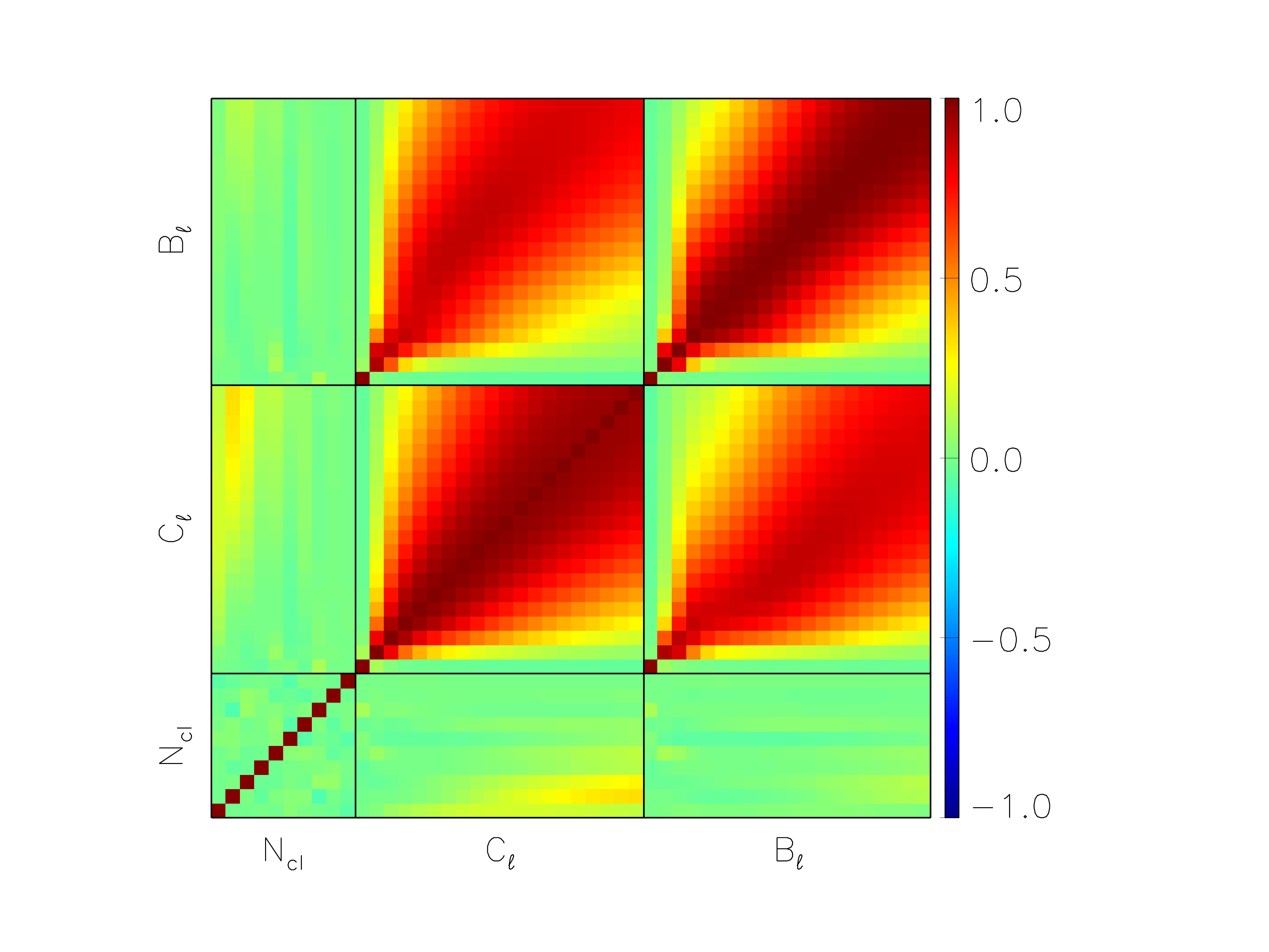}
\caption{Correlation matrix between galaxy cluster number count as a function of redshift $N_{\rm cl}$, tSZ angular power spectrum, $C_\ell$, and tSZ bispectrum, $B_\ell$, for the cosmic variance contribution to the uncertainties.}.
\label{ccor}
\end{center}
\end{figure}
\begin{figure*}[!th]
\begin{center}
\includegraphics[scale=0.45]{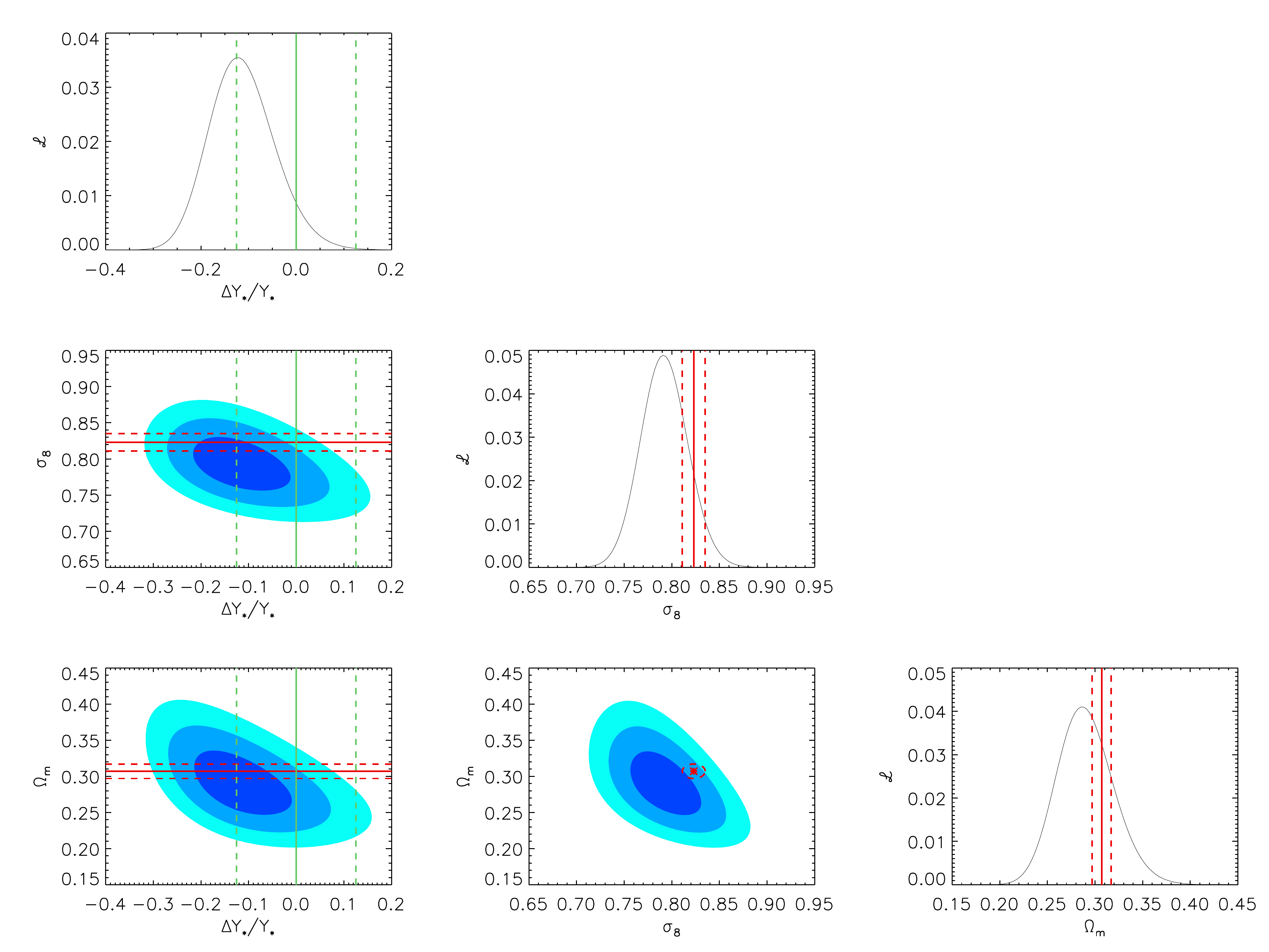}
\caption{Likelihood function on $Y_\star$, $\Omega_{\rm m}$, and $\sigma_8$ derived from the combined analysis of galaxy cluster number count, tSZ power spectrum, and tSZ bispectrum. Blue contours indicates the 1, 2, and 3 $\sigma$ confidence levels. The red solid lines-sample shows the constraints from CMB power spectrum, the dashed red lines shows the 1 $\sigma$ confidence level. The solid green line shows $(1-b) = 0.8$ and the dashed green lines show $(1-b)=0.7$ and $(1-b)=0.9$.}.
\label{fcosmo}
\end{center}
\end{figure*}

The covariance matrix of galaxy cluster number count, tSZ power spectrum, and tSZ bispectrum is particularly challenging to estimate. A complete analytic derivation would involve the computation of one to six points correlation functions.
Consequently, we estimated the covariance between probes by performing simulation of the galaxy cluster mass-function. 
The tSZ effect is sensitive to very high-mass galaxy cluster, thus we assumed that the mass-function covariance matrix is diagonal with respect to the galaxy cluster masses.

On Fig.~\ref{ccor}, we present the correlation matrix derived from the 100 simulations used already in Sect. \ref{secparbl}. Similarly to \citet{planckszc}, we consider 10 redshift bins from $z=0$ to 1 for galaxy cluster number count, $N_{\rm cl}$. For the tSZ power spectrum and bispectrum we consider multipoles from $\ell=0$ to 1000.
The cosmic variance contribution to the total uncertainties is dominated by non-gaussian terms that induces significant non-diagonal terms in the tSZ power spectrum and bispectrum covariance matrices.
We observe that the tSZ power spectrum is highly correlated with the tSZ bispectrum, with a correlation factor higher than $\simeq 0.8$ for the presented $\ell$ range.
Consequently, the combination of the tSZ power spectrum and bispectrum will not reduce significantly the cosmic variance contribution to the error on cosmological parameters.\\
We also observe that the galaxy cluster number count is not significantly correlated with the tSZ power spectrum and bispectrum.
The highest level of correlation, $\sim 50\%$, is obtained for low redshift galaxy cluster number count bins and the tSZ power spectrum at high $\ell$. The tSZ power spectrum and bispectrum at low $\ell$ are dominated by a small number of galaxy clusters, this explains the small correlation with number counts that considers several hundreds of galaxy clusters. At higher $\ell$ the tSZ power spectrum and bispectrum receives significant contribution from undetected galaxy clusters at higher redshift, inducing an overall small correlation level between number count and angular spectra.\\

We stress that this correlation matrix only represents the contribution from the cosmic variance to the total uncertainties. 
The total uncertainties also receive contributions from the instrumental noise and CIB residuals. To account for these additional uncertainties, we added random realizations of them to our 100 tSZ sky simulations.\\
We estimated the instrumental noise properties using Planck half-dataset difference for each frequency. We assumed that the noise in Planck frequency map is uncorrelated. 
Then, we propagated the noise through the MILCA linear combination.
Using this half-dataset noise maps, we estimated the noise inhomogeneities by computing the local standard deviation in a 4 degree FWHM gaussian beam. For the noise in the MILCA map, we also estimated the noise angular power-spectrum that can not be considered spatially uncorrelated due to the component separation process.
We performed 100 realistic simulations of the instrumental instrumental noise.\\
For the CIB component, we also performed 100 homogeneous correlated gaussian realizations using power spectra from the best fitting model from \citet{planckSZCIB} and propagate them through the MILCA linear combination.\\
We note that, given that these two sources of uncertainties are gaussian, they do not add correlations between the power spectrum and bispectrum measurements.

\subsection{Cosmological constraints}

Fog.~\ref{fcosmo}, presents the constraints we obtain on $Y_\star$, $\sigma_8$, and $\Omega_{\rm m}$. In that case we do not add prior on the $Y-M$ normalisation. And we interpreted the $Y-M$ calibration modification as an adjustement of $(1-b)$. 
We observe that the combined analysis favours a lower calibration for the $Y-M$ relation, leading to a best fitting value of $(1-b) = 0.71 \pm 0.07$. We also obtain $\sigma_8 = 0.79 \pm 0.02$ and $\Omega_{\rm m} = 0.29 \pm 0.02$.

We checked that individual results derived from each probe are consistent together as well as consistent with the total combination.\\
In order to investigate where the information is coming from, we examined the constraints from the three possible combination of two probes. We find that the combinaison of galaxy cluster number count and spectra allows to break degeneracies between the three considered parameters, giving optimal results on $\sigma_8$, $\sigma_8 = 0.79 \pm 0.02$, but information is still missing for the other parameters, $\Omega_m = 0.29 \pm 0.04$ and $(1-b) = 0.74 \pm 0.14$. The combination of the tSZ angular power-spectrum and bispectrum is efficient to break the degeneracy between $\Omega_m$ and $((1-b), \sigma_8)$, though still having a degeneracy between  $(1-b)$ and $\sigma_8)$ ; it gives $\Omega_m = 0.28 \pm 0.03$ and $\sigma_8 \left( (1-b)/0.7 \right)^{-0.42} = 0.81 \pm 0.02$. Finally the combination of number counts and the tSZ bispectrum gives $\Omega_m = 0.29 \pm 0.03$, $\sigma_8 = 0.79 \pm 0.03$, and $(1-b) = 0.70 \pm 0.07$. This last combination is thus particularly efficient to determine $(1-b)$ and is driving our constraint on the hydrostatic mass bias when combining the three probes.



\section{Conclusion and discussion}
\label{concl}

We have revisited the cosmological constraints derived from the tSZ angular power spectrum by performing a combined analysis of a tSZ $y$-map and Planck intensity map per frequency. 
This approach provided us with a measurement of the tSZ angular power spectrum robust with respect to the CIB contamination in the Planck tSZ $y$-maps, which was previously an important issue. From this analysis, we derived robust cosmological constraints, which come out consistent with previous works from the Planck collaboration. \\
We presented a halo-model description of the tSZ bispectrum and compared it with a measurement of the equilateral bispectrum of a Planck-derived tSZ $y$-map. Using the measurement we were able to set cosmological constraints, also robust with respect to contamination by near-Gaussian signals such as the CIB. These constraints come out both competitive and consistent with that from the tSZ angular power spectrum.\\
By computing their joint covariance, we demonstrated that the tSZ power spectrum and bispectrum present a high degree of correlation for the cosmic variance contribution to the uncertainties. However, the number counts of galaxy cluster, as performed in the analysis by \citet{planckszc}, are not significantly correlated with the tSZ power spectrum nor bispectrum.
Combining the number count analysis with our measurement of the tSZ angular power spectrum and bispectrum, we have been able to set tight constraints on the hydrostatic mass bias and cosmological parameter simultaneously.\\
The present results favor a value for the hydrostatic mass bias $(1-b) = 0.71 \pm 0.07$, consistent with the prior used in \citep{planckszc}, $(1-b) \in [0.7,1.0]$. It is particularly interesting to note that our combined analysis enables to break the degeneracy between cosmological parameters and the normalization of the scaling relation. It is the combination of the number counts and the bispectrum that drives the constraint on $(1-b)$, with the power spectrum helping to further reduce the cosmological error bars.\\
Finally, comparing these results with cosmological parameters derived from the Planck CMB analysis, we obtain an agreement between the two probes at 1 $\sigma$ level.

\begin{acknowledgements}
We acknowledge the support of the Agence Nationale de la Recherche through grant ANR-11-BS56-015.
\end{acknowledgements}

\bibliographystyle{aa}
\bibliography{sz_bispectre.bib}


\end{document}